\documentclass{PoS}
\usepackage{epsfig}
\usepackage{amsmath}
\usepackage{mcite}
\newcommand{\be}{\begin{equation}}
\newcommand{\ee}{\end{equation}}

\title{Hadron form factors using density-density correlators}

\ShortTitle{Hadron form factors using density-density correlators}

\author{C.~Alexandrou, \speaker{G.~Koutsou}\\
        Department of Physics, University of Cyprus, CY-1678 Nicosia, Cyprus\\
        E-mail: \email{koutsou@ucy.ac.cy}}
\author{H.~Neff\\
	}

\abstract{Gauge invariant density-density correlators yield detailed information on hadron structure.
Hadron deformation and form factors can be extracted for momentum transfers up to about $6$~GeV$^2$. We 
use stochastic techniques and dilution to compute the all to all propagator required for the exact evaluation 
of density-density correlators. We present first results for the pion form factor.
}

\FullConference{XXIVth International Symposium on Lattice Field Theory\\
                July 23-28, 2006\\
                Tucson, Arizona, USA}

\begin{document}

\section{Introduction}
A number of experiments have been performed recently to determine 
a possible deformation in the nucleon ~\cite{PRL:86-2963,*PRL:88-122001}.
Lattice QCD provides a model independent method for 
studying hadron deformation via the evaluation of density-density correlators. 
We show that it is now feasible  to evaluate density-density correlators for mesons and baryons
without any approximation. Such an evaluation requires a computation of all to all
propagators, which is carried out using stochastic techniques and dilution.
We show results for hadron charge distributions 
and demonstrate 
how hadron form factors can be extracted from density-density correlators.
We present first results for the pion form factor.
\section{Density-density correlators}
For general time insertions the density-density correlator is given by:
\be
C\left(\vec{y},t_1,t_2\right)=\int d^3xd^3z\left<h(\vec{z},t)\right|j_0^u(\vec{x}+\vec{y},t_2) j_0^d(\vec{x},t_1)\left|h(\vec{0},0)\right>
\label{Eq:ddcorr}
\ee
where $j^0$ is the time ordered density operator 
i.e. $j^0_f(x)=:\bar{q}_f(x)\gamma_0q_f(x):$ for a quark of flavor $f$, while $\left|h\right>$ denotes any
hadronic state. The integration over the $\vec{x}$ 
coordinate sets the sink momentum equal to that of the source. 
The integration over $\vec{z}$ sets the momentum
of the source and sink to zero.
These two sums require summation over 
both spatial coordinates of the quark propagator connecting the density insertions with the sink
and thus one needs to evaluate all to all propagators.

In this work we use stochastic and dilution techniques 
to compute all to all propagators.
It has been demonstrated \cite{PRD:58-034506} that an estimate 
for the all to all propagator can be computed 
by inverting the Dirac operator for an ensemble of noise vectors as sources.
If these noise sources are created in such a way that they obey:
\be
\left<\eta^a_\mu(x)\eta^{b\dagger}_\nu(y)\right>_r=\delta(x-y)\delta_{a,b}\delta_{\mu,\nu}\qquad\textrm{and}\qquad\left<\eta_\mu^a(x)\right>_r=0\label{Eq:noise_ensemble}
\ee
where $r$ denotes the number of noise vectors in the ensemble,
 then one can invert for each of these noise vectors 
and obtain an estimate for the all to all propagator:
\be
\left<\psi^b_\nu(x)\eta^{a\dagger}_\mu(y)\right>_r\rightarrow(M^{-1}(x,y))^{b,a}_{\nu,\mu}
\ee
where $\psi^b_\nu(x)=(M^{-1}(x,y))^{b,a}_{\nu,\mu}\eta^a_\mu(y)$ 
is the individual solution for each of the $r$ noise vectors. 
We additionally employ a method known as dilution, 
which gives a better estimate for the all to all propagator \cite{CPC:172-145}.
 The method essentially proposes a way for creating the noise 
source ensemble in Eq.~(\ref{Eq:noise_ensemble}). 
In this work we use even - odd, color - spin diluted noise vectors 
i.e. each of the $r$ noise vectors of the ensemble has random entries 
only on either odd or even spatial sites and on one 
color - spin component with all other entries set to zero. 
Since the 
time slices of the source, sink and density insertions 
are fixed the noise vectors are in effect already diluted in time.
\section{Wave functions}
We first consider the case where the two density insertions of Eq.~(\ref{Eq:ddcorr}) are taken at equal times.
This equal time correlator, shown schematically for mesons in Fig.~\ref{Fig:Meson}, reduces to the wave function squared in 
the non - relativistic limit. 
For baryons, the wave function is generally a function of two 
relative coordinates and thus requires the calculation of a three-density 
correlator depicted in Fig.~\ref{Fig:Baryon}. 
In this work we compute the one particle correlator 
with two density 
insertions on two different quark lines of the baryon. 
This is equivalent with integrating the three-density correlator 
over one relative coordinate as demonstrated in Ref.~\cite{PRD:66-094503} when the diagram in Fig.~\ref{Fig:Baryon}a was considered.

\begin{figure}[h]
  \begin{minipage}[t]{0.3\linewidth}
    \centering
    \scalebox{0.3}{\input{Meson.pstex_t}}
    \caption{The density-density correlator for a meson.}
    \label{Fig:Meson}
  \end{minipage}
  \hspace*{0.05\linewidth}
  \begin{minipage}[t]{0.6\linewidth}
      \parbox[t]{0.3\linewidth}{
        \centering
        \scalebox{0.3}{\input{BaryonA.pstex_t}}
      }
      \hspace*{0.2\linewidth}
      \parbox[t]{0.3\linewidth}{
        \centering
        \scalebox{0.3}{\input{BaryonB.pstex_t}}
      }
    \caption{The three-density correlator for a baryon. The density insertions are denoted by the crosses.}
    \label{Fig:Baryon}
  \end{minipage}
\end{figure}

In this study we use 200 unquenched gauge configurations 
produced by the SESAM collaboration~\cite{PRD:59-014509} on a lattice of size 
$16^3\times32$, at $\beta=5.6$ and $\kappa=0.157$. Using the mass of the nucleon in the chiral limit we obtain $a^{-1}=2.56$~GeV
yielding $m_\pi\simeq950$~GeV.
We use Wuppertal smearing to construct the interpolating fields for the source and sink. 
HYP smearing is applied to the gauge links that enter the 
Wuppertal smearing function. This decreases considerably the time 
interval needed for the suppression of excited states  
and allows to place the density insertions as early 
as three time slices from the source \cite{PRD:74-034508}. 
Although placing the density
insertions close to the source or sink is not important for the equal time
correlators it is crucial for the extraction of form factors
discussed in the next section (see discussion connected with Fig.~\ref{Fig:Pion_Platt}).
Similarly to satisfy the requirements for the extraction of form factors, the minimal
separation between source and sink that is required is $t=14$. We take the same source-sink time 
separation also in the evaluation of equal time correlators so that the same set of noise vectors can be used.
The evaluation of the density-density correlator
requires a forward propagator from the source  
and two all to all propagators, namely   
one from the density insertion at $t_1=3$ and one from the sink.
We note that the time slice of the second density insertion can 
be varied without additional cost a crucial observation for the evaluation of form factors from density correlators.
For each stochastic inversion we use 
6 sets of even - odd, color - spin diluted 
noise vectors. 
For comparison a separate computation was carried out 
where
no summation was performed on the sink 
and thus the momentum of the  hadronic state is not explicitly set to zero.

\subsection{Results}

In Fig.~\ref{Fig:WF_mesons} we show the density-density correlators 
for the pion and the $\rho$ meson. 
The distributions are compared to the case where 
no  explicit zero momentum projection is carried out.
If the non-relativistic limit is a good enough approximation so
that factorization of the center of mass momentum can be assumed then there
should be no difference
between the two evaluations.   
For the pion there is very little difference
showing that either  factorization 
is a good approximation
 or  exponential suppression of the time evolution sufficiently
damps out higher momentum states.
 On the other hand, the $\rho$ distribution is
broader when explicit projection to the state of zero momentum is carried out.
This shows that even for this large quark mass, factorization of the center of mass momentum does not
strictly hold.
These distributions are fitted using an exponential Ansatz
allowing the extraction of the root mean square (r.m.s.) radii. For the pion we find an r.m.s. 
radius $r_{r.m.s}=0.197(2)$~fm and for the $\rho$ meson 
$r_{r.m.s}=0.576(6)$~fm. The latter is  
approximately $25\%$ larger as compared
to what is obtained without explicit zero momentum projection.

\begin{figure}
  \vspace*{2.8cm}

  \begin{minipage}[t]{0.47\linewidth}
    \scalebox{0.55}{\input{Plots/wf_bin_mesons_k0.157_cmp.pslatex}}
    \caption{The correlators for the pion (bottom) and the $\rho$, in the zero-spin projection, (top). 
      The  crosses are the results of the exact evaluation and
      the  circles are the results when no explicit zero-momentum projection is carried out.}
    \label{Fig:WF_mesons}
  \end{minipage}
  \hspace*{0.03\linewidth}
  \begin{minipage}[t]{0.47\linewidth}
    \scalebox{0.55}{\input{Plots/wf_bin_baryons_k0.157_cmp.pslatex}}
    \caption[]{The correlators for the nucleon (top) and the $\Delta$ in the $\pm3/2$ spin projection
      (bottom).
      The notation is the same as that of  Fig.~\ref{Fig:WF_mesons}.}
    \label{Fig:WF_baryons}
  \end{minipage}
\end{figure}

In Fig.~\ref{Fig:WF_baryons} we make the same comparison for the nucleon
and the $\Delta$ correlators.
Both states are broader when all to all propagators 
are used to project to zero momentum. 
An Ansatz that describes these distributions 
accurately is of the form $e^{-(x/\sigma)^d}$. Fits to the 
distributions yield
$d\simeq1.2$ for the nucleon and $d\simeq1.5$ for the $\Delta$. 
The r.m.s. radius for the nucleon is only slightly bigger 
when the zero momentum projection is implemented
whereas for the $\Delta$ we find a
 $30\%$ increase.

\vspace*{-0.5cm}
\begin{figure}[h]
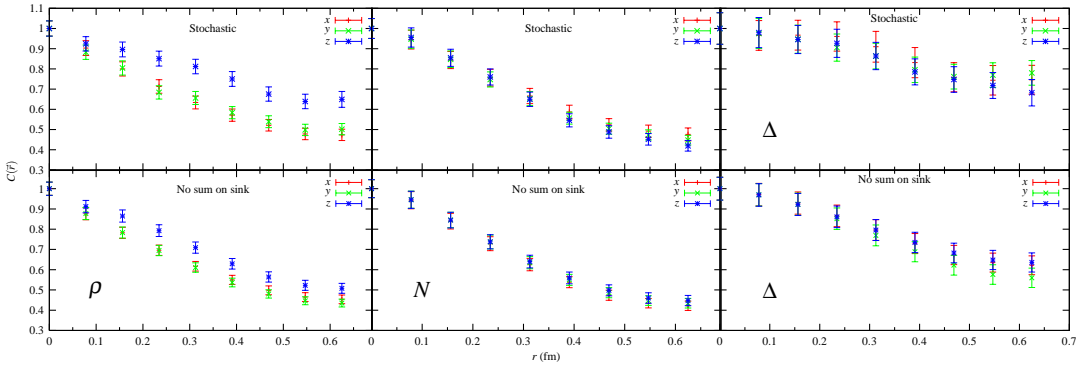

  \vspace*{2.5cm}

  \begin{minipage}[t]{0.3\linewidth}
    \scalebox{0.399}{\input{Plots/wf_axes_rho0_k0.157_cmp.pslatex}}
  \end{minipage}
  \begin{minipage}[t]{0.3\linewidth}
    \scalebox{0.399}{\input{Plots/wf_axes_nucleon1+2_k0.157_cmp.pslatex}}
  \end{minipage}
  \begin{minipage}[t]{0.3\linewidth}
    \scalebox{0.399}{\input{Plots/wf_axes_delta1.5_k0.157_cmp.pslatex}}
  \end{minipage}
  \caption{The $\rho$ (left), nucleon (center) and $\Delta$ (right) correlators projected along the axes. Top: with zero momentum projection. Bottom: with no
  zero momentum projection.}
  \label{Fig:axes_all}
\end{figure}

To probe hadron deformation
we project our results for the density-density correlators 
along the spin axis (taken as the $z$ axis) and perpendicular to the spin axis.  
In Fig.~\ref{Fig:axes_all} we show 
the projections for the $\rho$ meson in the zero spin projection, the nucleon and the $\Delta$ in the $\pm3/2$ spin projection.
We again compare the exact results
with the case where the summation on the sink is omitted.
The $\rho$ correlator shows an elongation along the spin axis. As expected
this elongation is much clearer when the state 
is projected to zero momentum. For the nucleon on the other hand,
 no deformation is detected 
within this method. 
The $\Delta$ baryon is too noisy to extract any useful information concerning a possible 
deformation. More statistics are clearly needed if one 
is to draw any definite conclusion for this hadronic state.

\section{Form factors}
The density-density correlator can be used to
 extract hadronic form factors \cite{PRD:43-2443}. 
Writing Eq.~(\ref{Eq:ddcorr}) on the lattice and 
inserting three complete sets of hadronic
states, one derives the expression:
\be
\sum_{\vec{p},n,s} \left|\left<\tilde{J}_s\left|h_s(0)\right.\right>\right|^2\frac{\left|\left<h_s(0)\right|j_0\left|n,\vec{p},s\right>\right|^2}{8M_h^2E_n(p)}e^{i\vec{p}\cdot\vec{y}}e^{-E_n(p)(t_2-t_1)}e^{-M_h(t-(t_2-t_1))},
\label{Eq:ff_halfway}
\ee
where $s$ represents internal degrees of freedom of the 
hadronic state and $\tilde{J}_s$ 
represents the smeared interpolating field used. In 
Eq.~(\ref{Eq:ff_halfway}) we assume 
isospin symmetry ($j_0=j_0^u=j_0^d$). 
We also assume that the time separation between the
density insertions from the source and sink
are large  enough so that excited state contributions are suppressed. 
Now if, in addition, $t_2-t_1$ is large enough and we take the  Fourier transform
over the $\vec{y}$ coordinate we arrive at the following expression:
\be
G^{hj^0j^0h}(s;\vec{q};t_1,t_2,t)=\sum_{s'} 
\left|\left<\tilde{J}_s\left|h_s(0)\right.\right>\right|^2
\frac{\left|\left<h_s(0)\right|j_0\left|h_{s'}(\vec{q})\right>\right|^2}
{8M_h^2E_h(q)}e^{-E_h(q)(t_2-t_1)}e^{-M_h(t-(t_2-t_1))},
\label{Eq:ffddcorr_general}
\ee
from where the hadron form factor can be extracted in the isospin limit. Note that in practice one can calculate 
the above quantity for any hadronic state 
with the same set of propagators. Additionally one has 
the form factor for all momentum transfers
since these follow from  the Fourier transform of the density-density
correlator.

To demonstrate that this method works in practice we consider
 the simple case 
of the pion. The pion as a pseudoscalar 
meson has only one form factor, which is given by:
\be
\left<\pi(p'_\mu)\right|J_\mu\left|\pi(p_\mu)\right>=(p_\mu+p'_\mu)F_\pi(q^2),
\label{Eq:pi_ff}
\ee
where $q_\mu=p'_\mu-p_\mu$. 
Combining Eq.~(\ref{Eq:pi_ff}) and Eq.~(\ref{Eq:ffddcorr_general}) 
the form factor $F_\pi(q^2)$ 
can be calculated by dividing the 
density-density correlator with an appropriate combination of two-point functions:
\be
F^2_\pi(q^2)=\lim_{\begin{subarray}{c}t-t_2\gg1,\\t_1\gg1,\\t_2-t_1\gg1\end{subarray}}\frac{4E_\pi(q)M_\pi}{(E_\pi(q)+M_\pi)^2}
\frac{G^{\pi j^0 j^0 \pi}(\vec{q};t_1,t_2,t)G^{\pi\pi}(0,t_2-t_1)}{G^{\pi\pi}(\vec{q},t_2-t_1)G^{\pi\pi}(0,t)}.
\label{Eq:pi_ff_unopt}
\ee
The ratio of two-point functions ($G^{\pi\pi}$) used in the
 above equation is the simplest that cancels the unknown overlaps of the interpolating fields with the pion state and the exponential time dependence.
However it is not the optimal as far as signal to noise is concerned. 
Ideally one wants to use two-point functions for the shortest possible
time separation
since the further the source is from the sink the noisier
 the propagator becomes. 
Thus one must find the optimal values for  the insertion times
and source-sink separation.
The dilemma is that although $t_2-t_1$ should be large enough to damp 
any intermediate excited states, the sink-source time
separation $t$ should be as small as possible 
so that the statistical noise on 
the two-point functions is minimal. Using smearing techniques we can ensure that ground state dominance holds for separations of the density insertions from 
the source and sink as small as three time slices.
This is demonstrated in Fig.~\ref{Fig:Pion_Platt} where we plot the 
density-density correlator for the pion with the 
sink separation fixed at $t=14$ and keeping the time
separation between the insertions at a constant value of $t_2-t_1=4$. 
We vary the time slice of the density insertions 
by one  starting from $t_1=3$ and ending at $t_1=7$.
It can be seen  that the results are identical for $t_2,t_1\ge 3$
and therefore we can set the first density insertion at $t_1=3$ and 
the second  at $t_2=11$ 
three time slices from the sink. 

\begin{figure}[t]
\vspace*{0cm}
\begin{minipage}[t]{0.47\linewidth}
\scalebox{0.575}{\input{Plots/wf_bin_pion_k0.157_platt.pslatex}}
\caption{The density-density correlator with the insertion time 
slices kept fixed at $t_2-t_1=4$.
 For this test we only used point to all propagators.}
\label{Fig:Pion_Platt}
\end{minipage}
\hspace{0.05\linewidth}
\begin{minipage}[t]{0.47\linewidth}
\scalebox{0.575}{\input{Plots/ff_pion_k0.157_cnvrg+VMD.pslatex}}
\caption{The pion form factor for various values of the time separation
between the density insertions. 
We also compare with the prediction using vector meson dominance (solid line).}
\label{Fig:Pion_Cnvrg}
\end{minipage}
\end{figure}

\begin{figure}[!h]
\vspace*{0cm}
\begin{center}
\begin{minipage}{0.975\linewidth}
\center
\scalebox{0.7}{\input{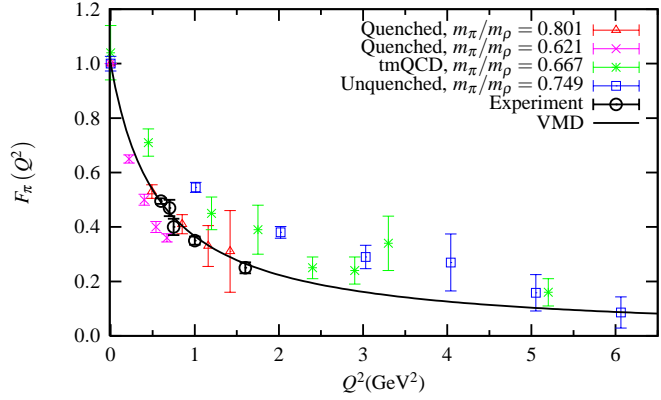}}
\label{Fig:Pion_FF}
\caption[]{The pion form factor versus $Q^2$ 
with similar $m_\pi/m_\rho$ ratios.
The triangles and crosses show
quenched results form Ref.~\cite{PRD:72-054506}, 
the  asterisks are results using twisted mass fermions~\cite{NPPS:140-299}, 
the squares are the results of this work 
and the circles denote the experimental results. 
The solid curve is the results obtained using
vector meson dominance with the physical $\rho$ meson mass.}
\end{minipage}
\end{center}
\end{figure}

Since $t_1=3$ and $t-t_2=3$ are the shortest separations we arrange so that the appropriate ratio involves two-point functions in terms of these time separations or 
time intervals of similar length e.g. $\frac{t_2-t_1}{2}$ which takes the maximal value of $4$. The optimal ratio is then given by:
\be
F^2_\pi(q^2)=\lim_{\begin{subarray}{c}t-t_2\gg1,\\t_1\gg1,\\t_2-t_1\gg1\end{subarray}}\frac{4E_\pi(q)M_\pi}{(E_\pi(q)+M_\pi)^2}\frac{G^{\pi j^0 j^0 \pi}(\vec{q};t_1,t_2,t) \left[G^{\pi\pi}_{SL}(\vec{q},t_1)\right]^4 G^{\pi\pi}_{SS}(0,t_1)}{G^{\pi\pi}_{SS}(0,2t_1) G^{\pi\pi}_{SS}(0,t-t_2)\left[G^{\pi\pi}_{SL}(\vec{q},2t_1)\right]^2 \left[G^{\pi\pi}_{SL}(\vec{q},\frac{t_2-t_1}{2})\right]^2},
\label{Eq:pi_ff_opt}
\ee
where $G^{\pi\pi}_{SS}$ and $G^{\pi\pi}_{SL}$ denote 
smeared - smeared and smeared - local pion two-point functions respectively. 
To check for convergence we plot in Fig.~\ref{Fig:Pion_Cnvrg} $F^2_\pi(q^2)$, given in Eq.~(\ref{Eq:pi_ff_opt}), 
as a function of $t_2-t_1$.
As can be seen the pion form factor has converged for $t_2-t_1=8$,
which is the maximum even separation that can be achieved on this 
lattice. On the same plot we compare the form factor to the results obtained
assuming vector meson dominance, i.e. taking $F_\pi(Q^2)=\frac{1}{1+Q^2/m^2_\rho}$
where $Q^2=-q^2$ is the Euclidean momentum transfer squared. For $m_\rho$ we take the $\rho$ meson mass computed on the lattice which
is $m_\rho=1.270(8)$~GeV. In Fig.~\ref{Fig:Pion_FF} we compare 
the pion form factor to other recent lattice results
extracted using the standard approach of evaluating three-point functions.

Within this method we obtain reliable results up to momentum transfers of 
$Q^2\simeq6$~Gev$^2$. The momentum transfers we can 
extract are limited by the fact that for high momentum transfers 
the density-density correlator given in Eq.~(\ref{Eq:ffddcorr_general}) becomes too noisy, 
becoming negative
and the square root cannot be taken.

\section{Conclusions}

Stochastic techniques combined with dilution are employed in the evaluation
of all to all propagators needed for the exact computation 
of density-density correlators. 
The equal time four-point correlators reduce to the wave function 
squared in the non-relativistic limit and thus provide
detailed information on the quark distributions inside the hadrons.
In particular the deformation of the hadron can be studied.
We clearly detect a deformation in the case of the $\rho$ meson.
However, no definite conclusion regarding deformation
can be reached for the $\Delta$, at least within these statistics. 
Another application of density-density correlators is the 
evaluation of hadron form factors. We apply this method
to compute the pion form factor up to $Q^2\sim 6$~GeV$^2$. 
The results we obtain
within  this method are comparable to the results extracted from the 
standard approach using three-point functions.  

\bibliography{latt06.bib}

\end{document}